# Compaction Dynamics Of Metallic Nano-Foams:

# A Molecular Dynamics Simulation Study

Farid F. Abraham[*] and M. A. Duchaineau

Lawrence Livermore National Laboratory

Livermore, California

**ABSTRACT**

We investigate, by molecular dynamics simulation, the generic features associated with the dynamic compaction of metallic nano-foams at very high strain rates. A universal feature of the dynamic compaction process is revealed as composed of two distinct regions: a growing crushed region and a leading fluid precursor. The crushed region has a density lower than the solid material and gradually grows thicker in time by "snowplowing." The trapped fluid precursor is created by ablation and/or melting of the foam filaments and the subsequent confinement of the hot atoms in a region comparable to the filament length of the foam. Quantitative characterization of nano-foam compaction dynamics is presented and the compacted form equation-of-state is discussed. We argue that high-energy foam crushing is not a shock phenomenon even though both share the snowplow feature.

**INTRODUCTION**

There is a premium in predicting the thermodynamic and mechanical response of materials subjected to shock-induced and shockless compression, either with mechanical impactors (gas gun) or laser driven ablators. The emergence of high-energy laser systems, such as the National Ignition Facility (NIF), presents the opportunity to create and interrogate states of matter under unprecedented extreme conditions of pressure, temperature, and strain rates. NIF experiments will also be used to conduct astrophysics and basic science research and to develop carbon-free, limitless fusion energy. In the NIF Energy program, a metallic nanofoam shell has been proposed as an important ingredient in the design of the double-shell

Inertial Confinement Fusion (ICF) target. The dynamics of its collapse upon implosion will play an essential role in achieving inertial confinement fusion. However, what is lacking is a fundamental understanding of the relationship between the microphysics of the materials properties of the drive impactor ("reservoir") and the target materials. An atomic level simulation effort can serve as the basis to relate the microstructure of the reservoir impactor to the pressure profile on the target. The results of these ultra-scale atomic level simulations can be significant in improving the predictive capabilities of continuum level simulations shock compression NIF experiments.

The study of matter under planetary interior conditions of high pressure and relatively low temperature environments is an important research area for NIF (Lee et al. 2004). In the traditional laser-based EOS experiment, a strong shock is launched in a material which instantaneously increases both the temperature and pressure tracing the Hugoniot line in the phase diagram. However, broader knowledge of the equation-of-state is required. Recent work by Remington and co-workers (Smith et al. 2007, Lorenz et al. 2005, Edwards et al. 2004) has demonstrated the possibility of a quasi-isentropic compression on a laser based platform. The intense laser pulse drives a reservoir which unloads across a gap onto the sample of interest. Low density porous materials are used as a reservoir because they provide a convenient method to independently vary density without changing the atomic number. The basic idea is to vaporize the foam, and compress the hot vapor, thereby increasing the pressure and temperature of the "vapor piston" smoothly. Molecular dynamics simulation techniques can provide design tools to tailor the pressure pulse for isentropic compression of a solid.

In this study, our simulation systems are composite crystal/nanofoam sandwiches. We have constructed physical models and developed the simulation and analysis programs with supporting visualization tools for application to study high strain rate compaction of metallic nano-foam by large-scale computation. Our study incorporates computer-generated nanofoams created by rapidly quenching a high temperature, phase-separating fluid. Our computer foam has been used earlier to study surface-stress-induced relaxation of Au nanostructures (Biener et al. 1980). This present study elucidates the generic features associated with the dynamic compaction of metallic nano-foams at very high strain rates.  Quantitative characterization of nanofoam compaction dynamics is presented and equation-of-state of the compacted form is discussed.

**COMPUTER MODELLING FOR THE SIMULATION**

Our simulation method is Molecular Dynamics (MD) (Abraham 1986, Allen & Tildesley, 1987), where it is assumed that the motion of the atoms is govern by Newton's 2nd law. The second ingredient is the chosen interatomic potential. It can be simple or it can be complex. While choosing a complex potential to describe a particular material is often desired, the physics of a complicated process can be more transparent and discovery of "generic" behavior may be more readily forthcoming with the choice of a simple potential.  We assume the Voter-Chen EAM potential (Voter 1993) for copper as our simple potential. We acknowledge the limitations associated with this choice of potential and will discuss some, such as the neglect of ionization at very high temperatures and pressures. However, we emphasize that our goal is to elucidate the generic features of the compaction dynamics of nanofoams. Quantitative predictions go beyond our current goal of elucidating generic features.

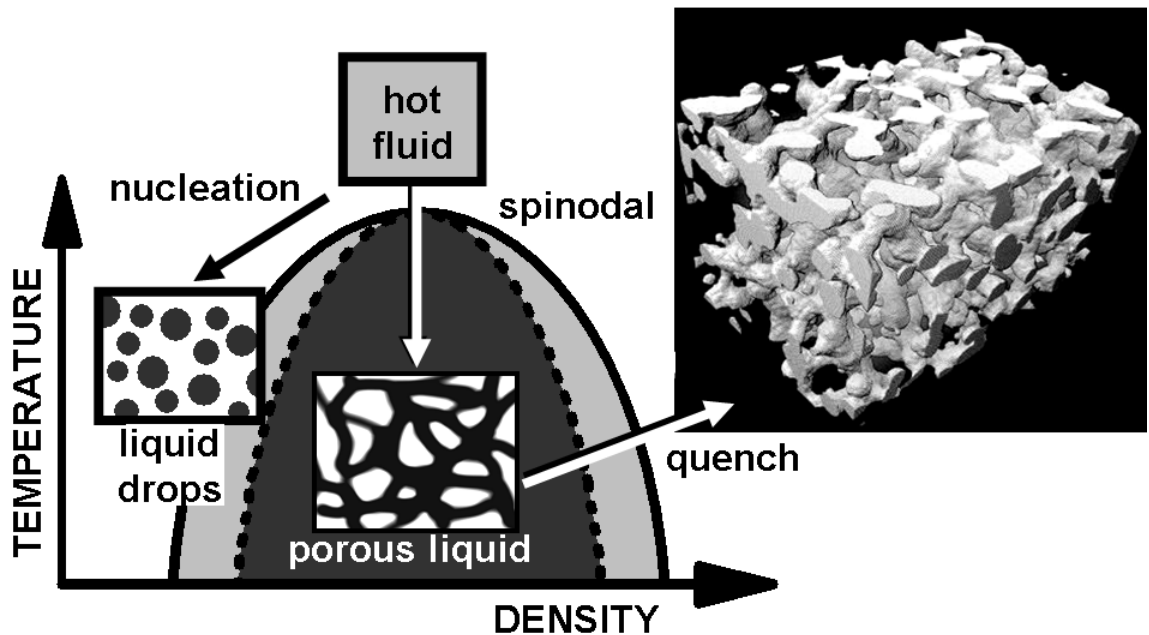

FIG. 1: Making computer foam by spinodal decomposition. The method is the rapid quenching of a superheat fluid into the two-phase region followed by the freezing in early time of the phase-separating liquid.

Making computer foam is the first step. We take the material, copper in this study, to its super critical state and rapidly quench it to the middle of its two-phase fluid region of its phase diagram. It begins to undergo phase separation by spinodal decomposition (Abraham, 1979). We stop the fluid phase separation process by rapidly quenching to the solid region, giving rise to nanoporous foam (FIG. 1). The initial porous sample is small and the filaments are irregular and amorphous.

For studies of the material response of pore structures under various conditions, it is desirable to construct pore structures at larger length scales, with controlled atomic arrangements (e.g. perfect crystal, grains, etc), and with specified density profiles or filament cross-sections. We briefly outline our topological filtering approach to analyze and synthesize such designed pore structures using the original small, amorphous sample as a template. Overall the analysis and synthetic processing involves six steps. These six steps are illustrated in Figure 2. a) Input an initial set of atom positions from the spinodal decomposition process; b) Make a proximity field of original atoms on a regular grid; c) Compute a signed distance field relative to the solid/void interface surface; d) Perform a topology preserving surface propagation from the interface surface to produce a topologically "clean" distance field and curved skeleton of the pore filament structure; e) Compute a distance field from the curved skeleton, re-scaled to produce uniform density profiles; f) Use the re-scaled distance field to "carve out" atoms with identical pore filament topology as the original sample, but with specified scales, density profiles, and atomic arrangements (Gyulassy et al. 2007, Laney et al. 2002). Details are to be published (Duchaneau and Abraham, 2010). A movie of the nano-foam structure may be viewed at:

http://www.llnl.gov/largevis/atoms/challenge2007

For the three-dimensional pore structure used in this study, a solid/void interface was determined (shown as a green surface in the movie), along with the solid filament centerlines (shown in red). We have created foams with filament sizes up to 35 nanometers while experiments are creating foams between 20 & 100 nm.

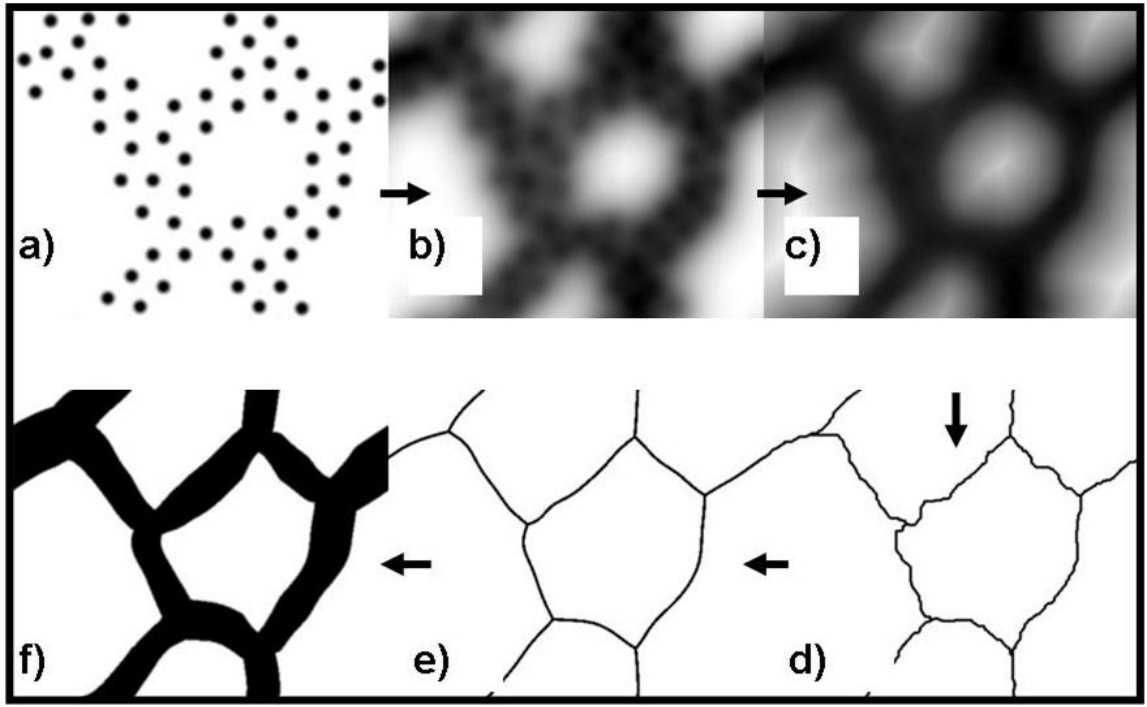

FIG. 2: Graphically outlining the topological filtering approach for analyzing and synthesizing nano-pore structures. See text for explanation.

We used this computer foam to construct the system for the shock simulations. We call it a "sandwich," and its building blocks are comprised of foam, crystal and vacuum constructs. We will study the "sandwich configuration" in Figure 3.

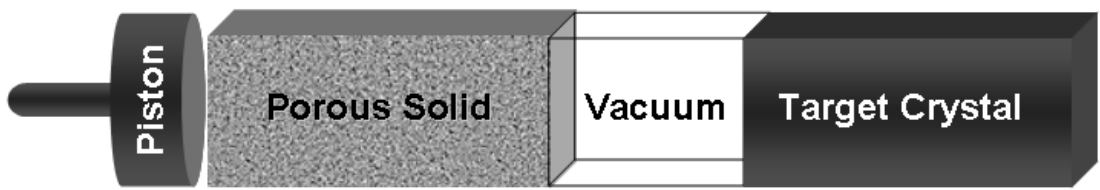

FIG. 3: The simulated sandwich configuration. Its building blocks are comprised of foam, crystal and vacuum constructs.

The purpose of the nano-porous foam block is to provide a source for creating hot vapor by high-energy impact. The anticipated history can be easily summarized. The piston moves at constant velocity to the right, crushing and heating the foam block by a rapidly moving front. We describe this as a "compaction" front in contrast to the popular "shock" front description. It is anticipated that if hit hard enough, the foam completely vaporizes. Hot vapor expands from the heated foam into the vacuum and is compressed by the driving piston. This, in turn, creates a continually rising pressure on the face of the right-hand target crystal. In this paper, our primary interest is the qualitative features of the equation of state of the compacting foam. The computer foam in this study is taken to be 15% of solid copper. The target dynamics will be treated in a future study.

## THE SIMULATIONS

### Universal Dynamical Features Of A Shocked Sandwich

The universal feature of the compaction process traveling through the porous solid has emerged from our simulations. It is made up of two regions: the "growing compacted region" and the "trapped fluid precursor." (see Figure 4a).

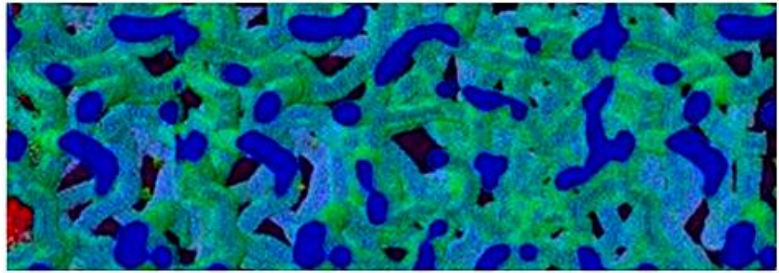
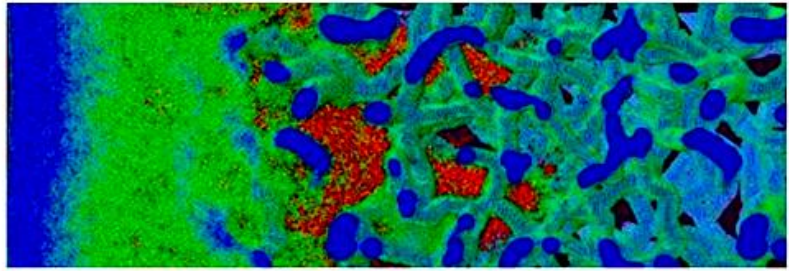

FIG. 4a: Two distinct regions defining the universal feature of the compaction process. The compacted foam and the vapor precursor are shown. The colors are associated with the potential energy of the individual atoms, blue being the energy of a solid copper atom and red being a weakly bonded or isolated atom.

In contrast to a shocked solid which achieves constant density immediate behind a sharp front, foam compaction is an evolving phenomenon where growing densification occurs as the region falls back from the forward moving broad front (see FIG. 4b). Furthermore, the compacted region has a peak density lower than its cold solid material. A feature shared by shocking and by compacting is that growth is achieved by snowplowing.

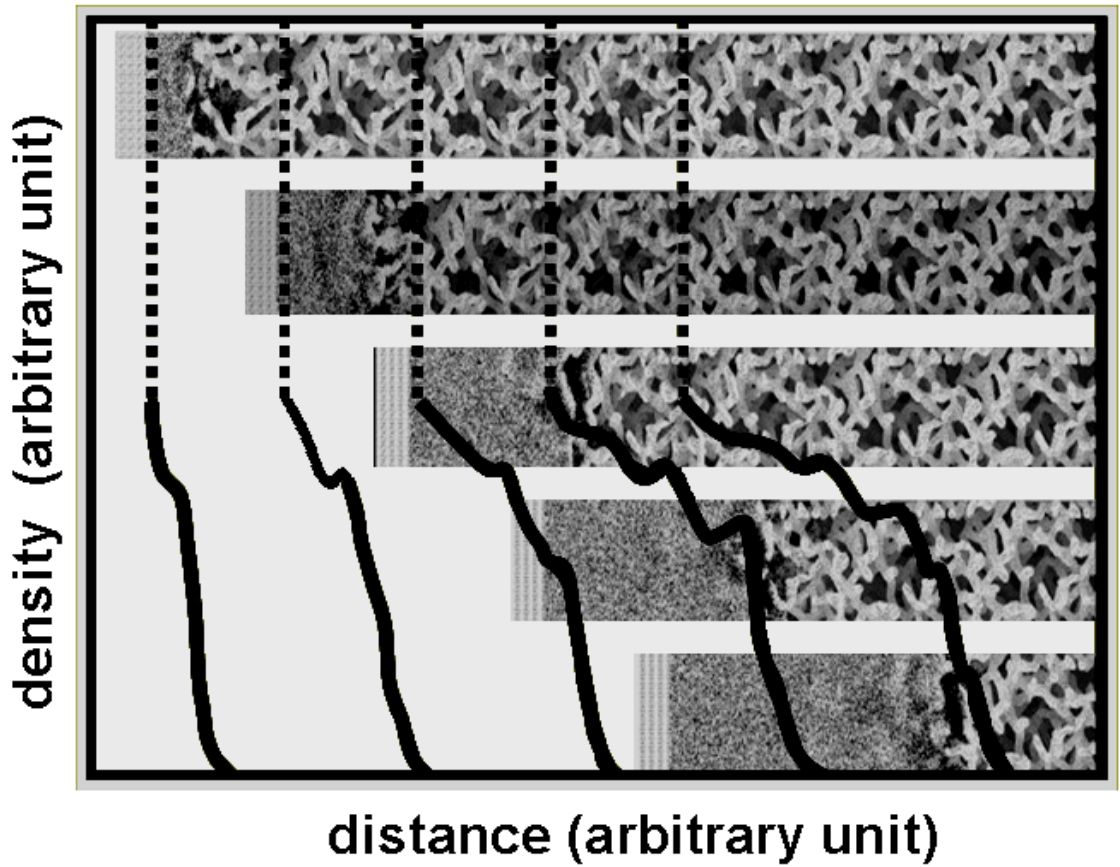

FIG. 4b: Growing densification of the crushed foam as a consequence of snowplowing. The vertical dash lines define the piston position and the solid lines are the respective instantaneous density profiles. Superimposed in the background are snapshots of the sandwiches.

The trapped fluid precursor is created by ablation and/or melting of the foam filaments and the subsequent trapping of the hot atoms within a "mean-free-path" comparable to the filament length of the foam. That is to say, the high-energy fluid atoms do not experience free expansion. Instead, they traverse a maze of uncorrelated random channels restricting free flow to the opened vacuum. This fluid precursor can be liquid or vapor, depending on the piston impact speed (Fig. 5). It rapidly achieves an approximate fixed length, traveling at a speed comparable to the compaction front. The compaction front speed is greater than the piston speed but slower than the shock speed in the perfect solid. This picture is consistent with a recent experiment (Dittrich, 2009) where a calorimeter measurement suggested that a significant temperature rise preceded the compaction front traveling through the foam.

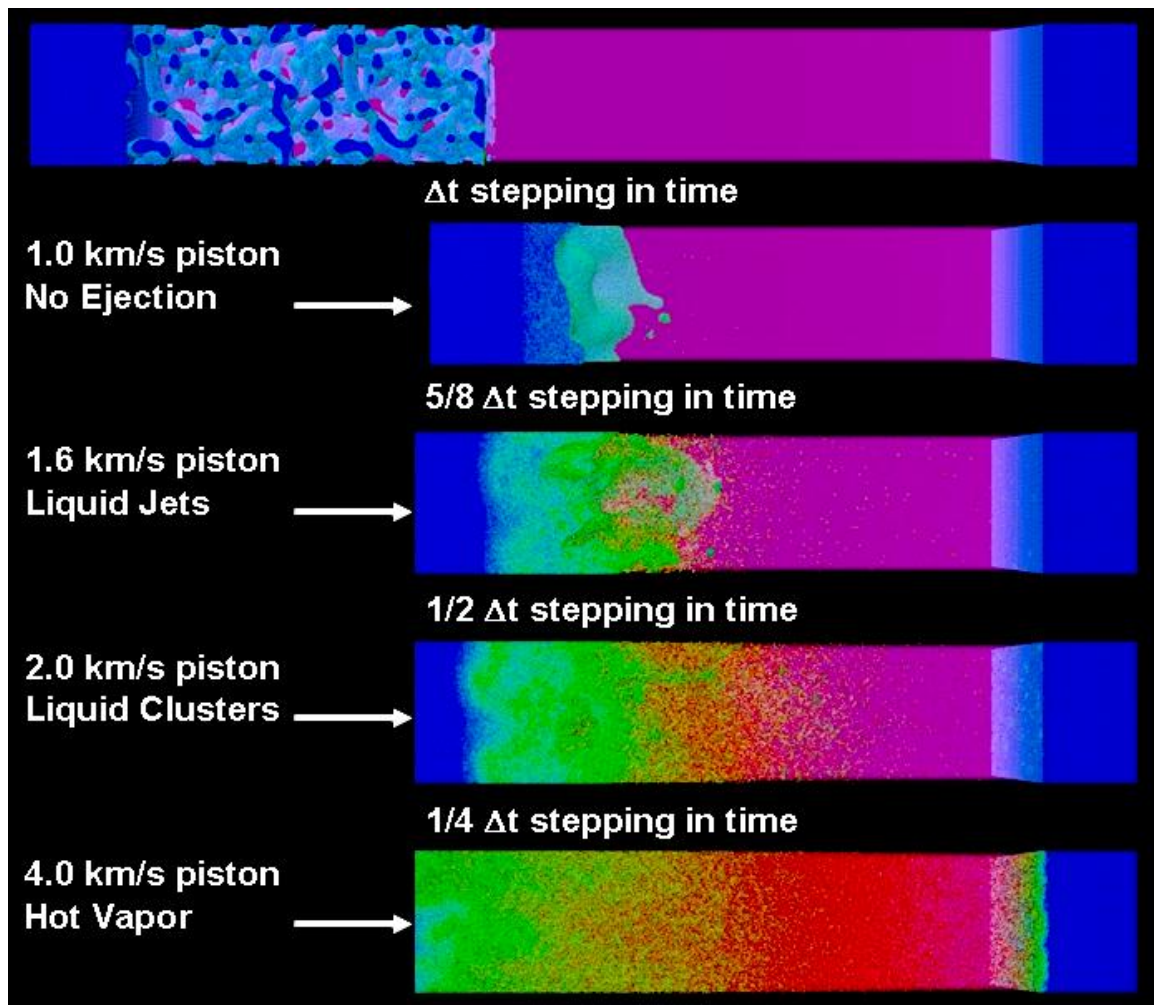

FIG. 5: The compaction precursor breaking out into the vacuum. This fluid precursor can be liquid or vapor, depending on the piston impact speed. See caption of Figure 4a for description of colors.

**Features of the Ejecta Leaving the Foam**

We drive our sandwich with pistons traveling at speeds of interest to recent experiments and have discovered a "phase change" associated with the ejecta at the slower speeds. This is best seen as the ejecta leaves the foam boundary (FIG. 5). At speeds below 1 km/s, there was only foam compaction and no ejecta. As the piston speed increased from 1 km/s to 4 km/s, we observed a liquid front, passing to liquid jetting, liquid cluster spraying, and, at the highest speed, hot vapor ejecta. Since the recent EOS experiments are presently approaching 20 km/s, we can conclude that the ejecta will be a hot vapor at these higher piston speeds. For Figure 5, we have normalized the individual snapshot times so that the apparent rates of collapse of the four sandwiches are the same. In Figure 6, we see a snapshot of dramatic liquid ejecta at a piston speed of 1.6 km/s.

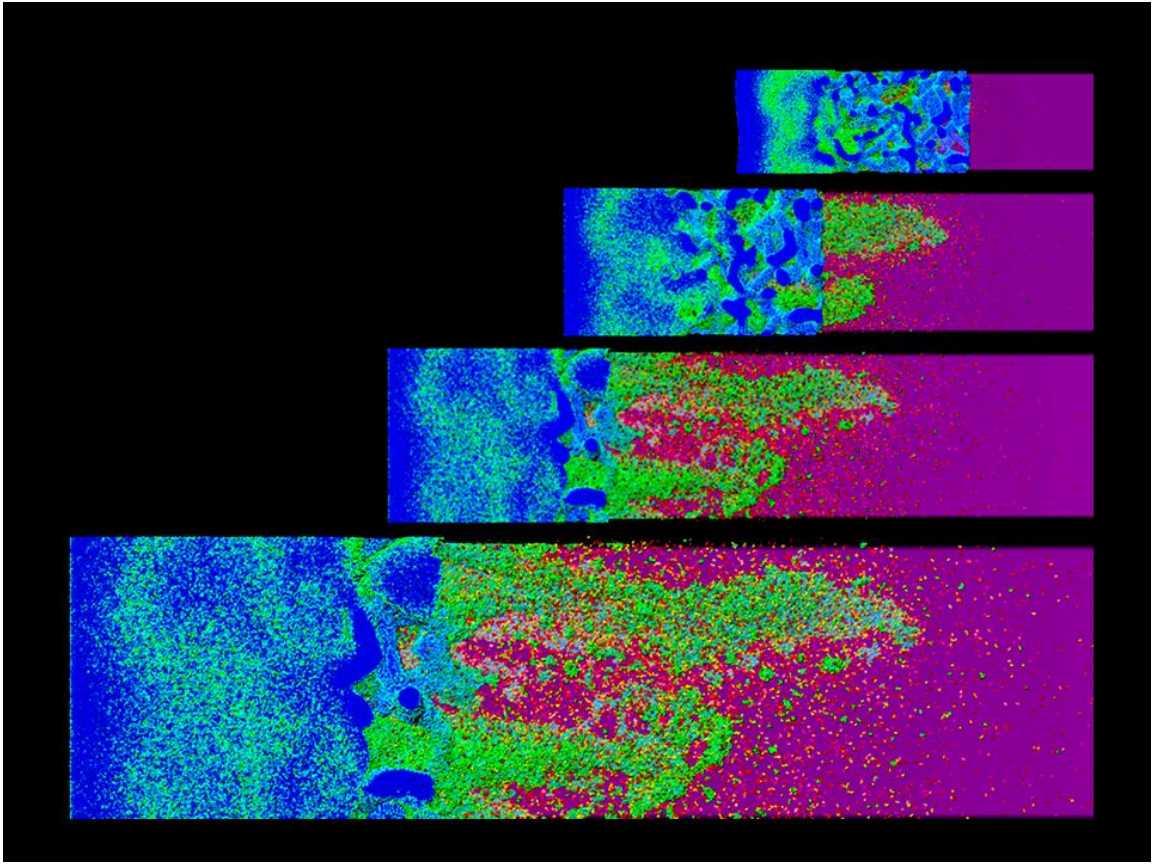

FIG. 6: Snapshot of liquid ejecta at a piston speed of 1.6 km/s in grey scale. See caption of Figure 4a for description of colors.

**Equation Of State Simulations Of The Compacting Foam**

We have extended our equation-of-state study for the compacted form by simulating longer foam slabs and higher piston speeds. The computer foam in this study is taken to be 15% of solid copper. Figure 7 shows the respective density profiles for piston speeds 1 to 20 km/s at decreasing snapshot times so as to capture the same piston location. The chosen piston location is where the compacted regions are approaching their greatest thickness; i.e., the respective fronts are near breakout or the foam-vacuum interface. We note that the density profiles look essentially identical for a piston speed equal to and greater than 5 km/s. There is not perfect overlap of the steep front profiles, but the variation is not systematic in piston speed. The commonality of the different profiles is striking and quite unexpected.

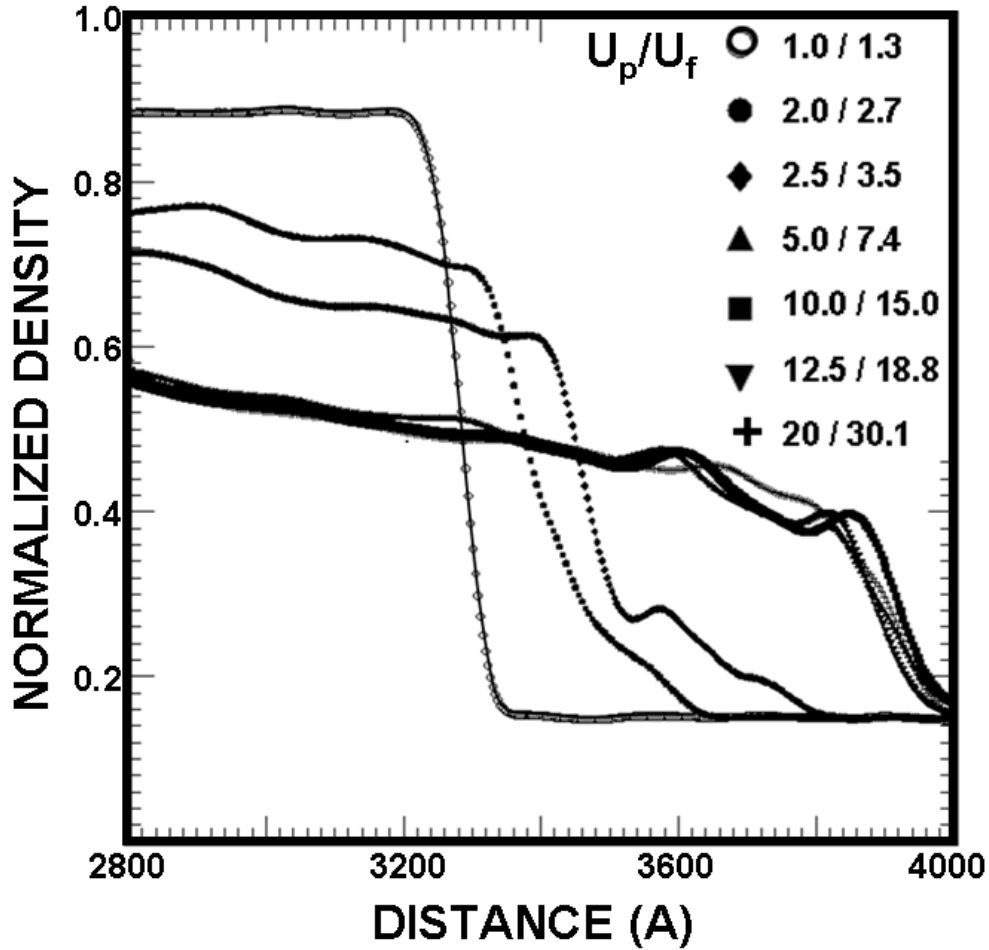

FIG. 7a: Density profiles for piston speeds 1 to 20 km/s, respectively. The snapshot times are chosen so as to capture the same piston location near breakout.

Extending the 5 km/s simulation to four times in time and distance (FIG. 7b), we note that the smoothed profile of the compacted foam maintains its shallow slope with the contact density at the piston position not changing (dashed line).

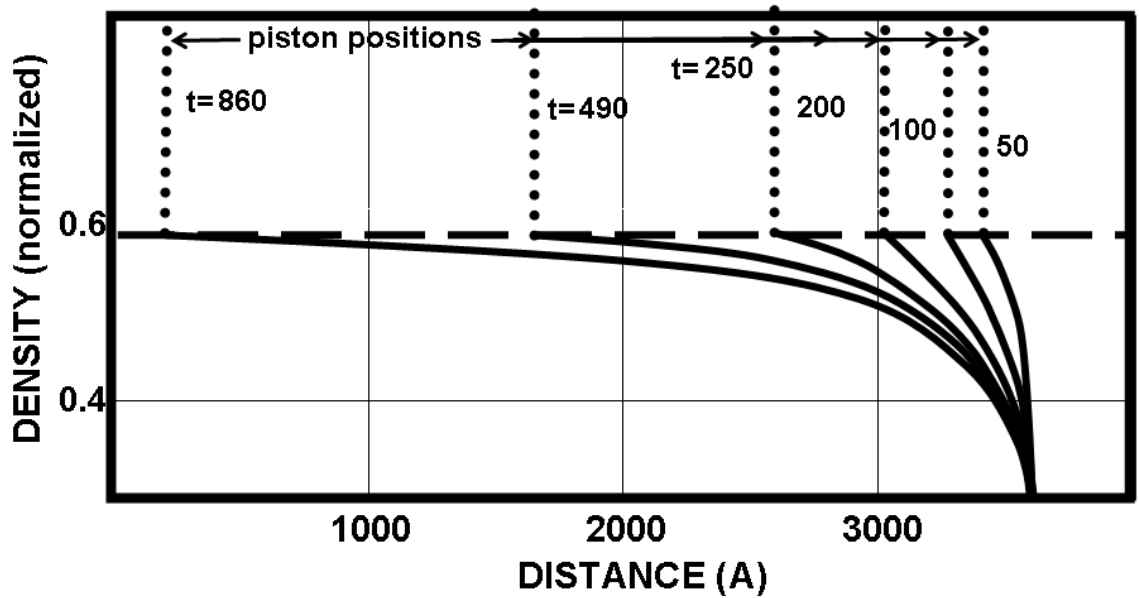

FIG. 7b: Smoothed density profiles at increasing times, t, for 5 km/s simulation. Profiles are centered with respect to the compaction front position and visually smoothed to demonstrate the form change. Time is in arbitrary units.

Mass and momentum conservation applied to the simple "snowplow model" (Zel'dovich & Raizer 2002) can explain the foam densification of the compaction front. Conservation of mass is the essential ingredient for predicting the front speed. There is good agreement between the simulated speeds and the snowplow model; i.e., the jump condition prediction of Equation 1 (see FIG. 8). Our extended simulations agree with very recent (but limited) experiment data (Page 2009).

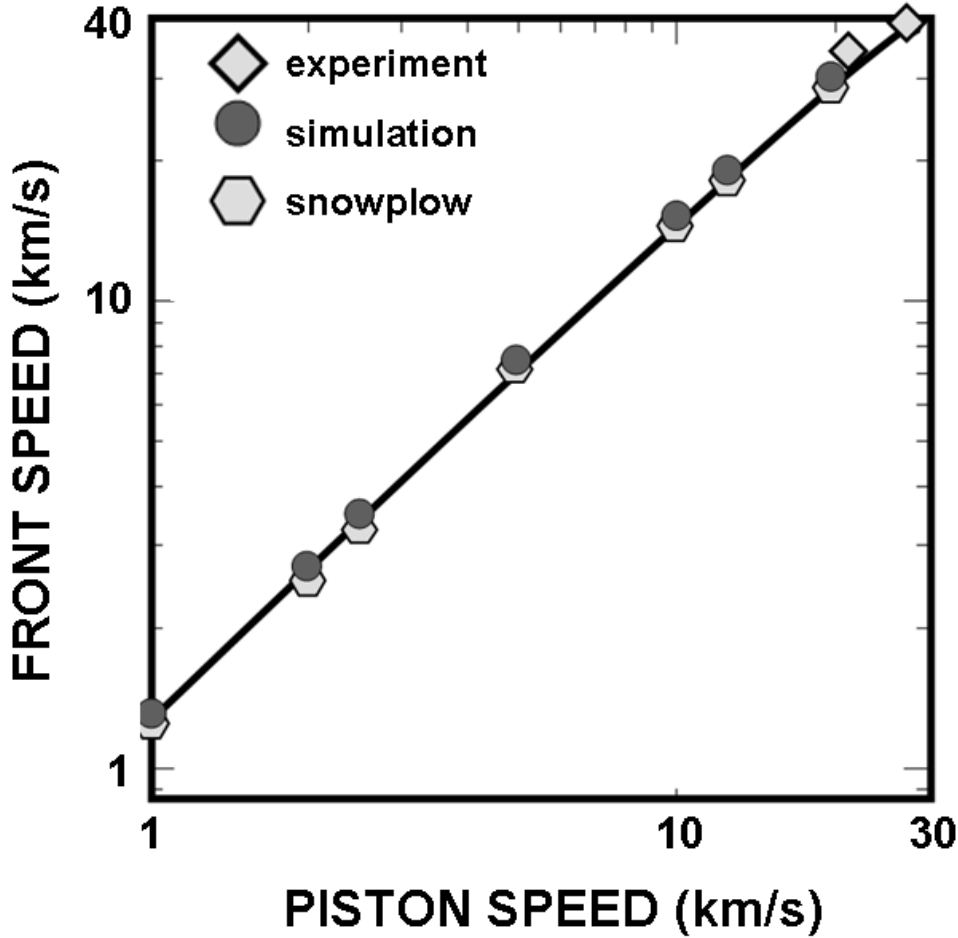

FIG. 8: Comparison of the front speeds as a function of the piston speeds for the simulations, snowplow "jump" predictions and experiment. The agreement is good.

There is a very important point to make concerning this success. One might be surprised that we have such good agreement with experiment; e.g., our simple model assumes "billiard ball" atoms that remain neutral irrespective of temperature and pressure. Global conservation of mass yields the simple relation

$$U_f/U_p = 1/[1-(\rho_0/\rho_c)] \qquad (1)$$

where $\rho_0$, $\rho_c$, $U_f$, $U_p$ are non-compacted foam density, compacted foam density, front speed and piston speed, respectively. We see that this expression has a weak dependence on the ratio of "before and after compaction" foam densities and, as such, does not place a high demand on modeling accuracy. This is a consequence of the snowplow phenomenon. The compacted regions do not have a strictly flat (constant) density, but assuming an average density is sufficient to obtain excellent agreement. While this agreement is not a sensitive measure of our knowledge of the compaction dynamics, this insensivity plays an important role in estimating ionization effects discussed later in the paper.

Similarly, the atomic virial pressure of Kirkwood and the snowplow estimate,

$$P = \rho_0 \, U_f \, U_p \qquad (2)$$

are in agreement (Fig. 9). This is a consequence of momentum conservation across the front, a global conservation requirement that is necessarily satisfied. The snowplow pressure P is calculated using the measured front speed from simulation. Again, we realize that such agreement is not a sensitive measure of our knowledge of the compaction equation-of-state by simply examining Eq. 2.

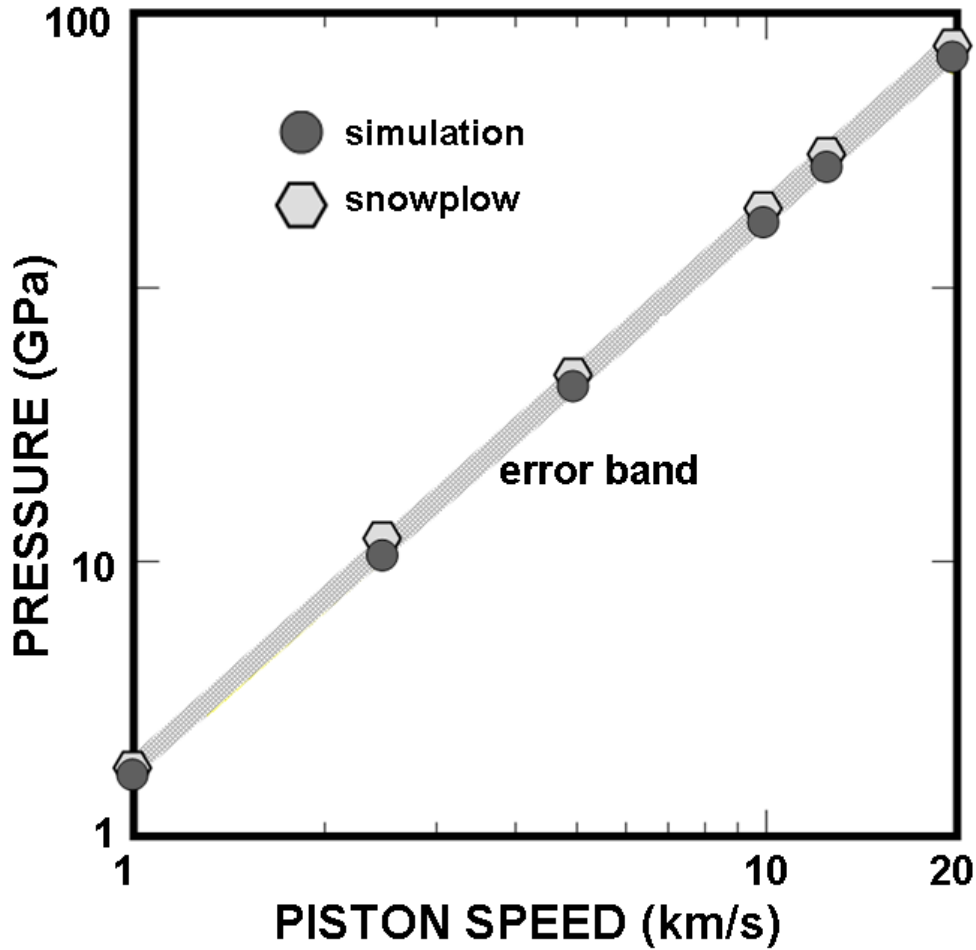

FIG. 9: Comparison of pressures as a function of piston speeds for the simulations and the snowplow "jump" predictions. The error band is estimated using the simulation uncertainly in the virial pressure. The comparison shows good agreement.

We mention an important point in calculating the pressure from the virial equation. Care must be taken to use the local density in this expression for calculating the pressure profile. Using the local density, one gets the correct virial pressure which is constant throughout the compacted foam (even though the density is not) and equal to the pressure estimated from the conservation of momentum requirement.

The equation of state EOS of the compacted copper foam is presented in Figure 10. The figure shows "non-ionizing" and ionizing states. The applicability of the "non-ionizing" EOS for real copper foam is in doubt much beyond a piston speed of 10 km/s because of ionization of the copper atoms. To estimate the effect of ionization on the EOS, we consider two different methods. The first method uses tabulated equation-of-state data for hot-dense matter based on sophisticated model calculations and experiment (More et al. 1988, Rozsnyai et al. 2001). In our earlier discussion of the "snowplow" model, we emphasized that global conservation of mass and momentum leading to Eqs. 1 & 2 demonstrates a weak density dependence of front speed, as well as the pressure. Furthermore, the front speed found from simulation agrees with the experimental measurements. This leads us to assume that the density and pressure for a given simulation is the same for both the neutral and ionized state, and we employed these tables to determine the temperatures of the ionized state. The results are shown in Figure 10 as "del-points." We note that the ionized state temperature at 20 km/s is a factor of four lower than the MD result.

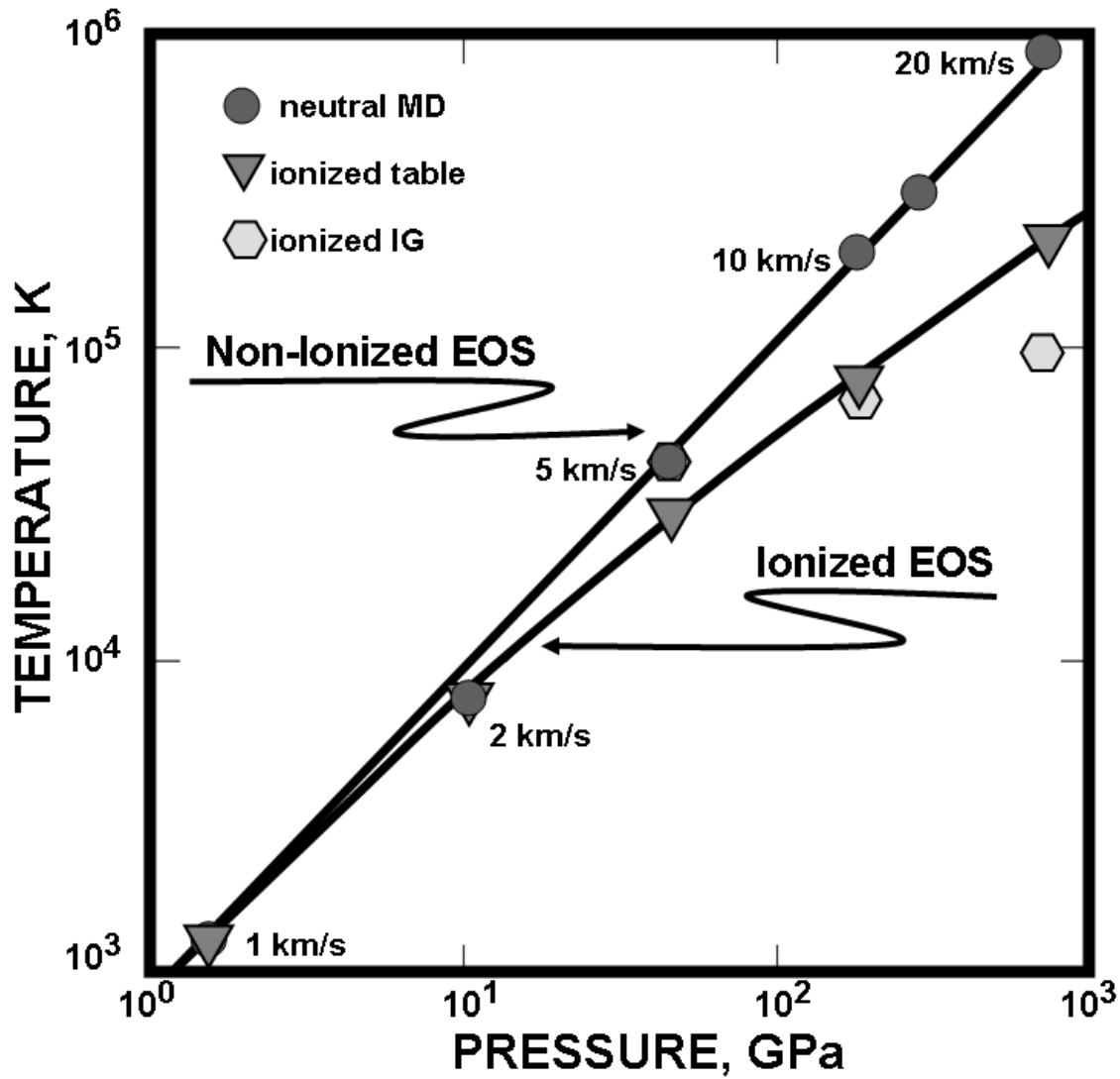

FIG. 10: Equation of state of compacted copper foam from the molecular dynamics simulations (non-ionized EOS) and from the approximated methods for correcting the MD results (ionized EOS). The lines are fitted and are included to guide the eye.

For the second method, we invoke conservation of energy in the ideal vapor (IG) approximation ( Zel'dovich & Raizer 2002, 192-193pp). We discuss this IG approach even though the table lookup is far more accurate since it may be useful to the reader for pedagogical reasons. We denote as state 0, the crushed foam where the copper atoms are constrained to be neutral. This is our MD case. Energy, number of atoms N, average number of electrons ionized per atom are denoted by E, N, Z, respectively. In the ideal vapor approximation, the energy is

$E_0 = 1.5 N T_0$

for the neutral system and

$E_i = 1.5 N (1+Z_i) T_i + N\varepsilon$

for the ionized system where $\varepsilon$ = average energy used to free Zi electrons from an atom. Assuming the energy and the number of atoms are conserved and that the density is constant, we find then the following relation:

$T_0 = (1+Z_i) T_i + (\varepsilon / 1.5)$ (3)

The ionization energies of copper are 0 to +1: 745.5 kJ/mol, +1 to +2: 1958 kJ/mol, +2 to +3: 3554 kJ/mol, +3 to +4: 5326 kJ/mol. Using the Eq. (3) between the two temperatures, we go up the ionization scale until it predict the lowest positive ionized temperature. In Fig. 10, the ideal gas (IG) predictions are plotted as hexagons. At the highest pressure, the IGA ionization lowers the temperature by an order-of-magnitude. This is a very simplistic model is pedagogical and shows explicitly why there would be lower temperature due to ionization. The first estimate using the accurate tables is favored.

In Figure 11, pressure-density EOS is compared to the known copper Hugoniot (More et al. 1988, Rozsnyai et al. 2001) . We note that the compacted foam state is not accurately represented by the shock hugoniot of low density copper foam. This is not surprising from all that we have learned earlier. One must be cautious in attempting to characterize the high-energy crushing of foam as a shock phenomenon. However, this does not preclude the very good agreement of front speed versus piston speed between the MD simulations and the copper foam hugoniot. Comparison of front speed versus piston speed for molecular dynamics simulations (MD), experiment and the copper foam hugoniot is shown in Figure 12. The agreement between all three is good, though MD and experiment is best. This reinforces the observation that the dependence of front speed as a function of piston speed is an insensitive measure of the equation-of-state.

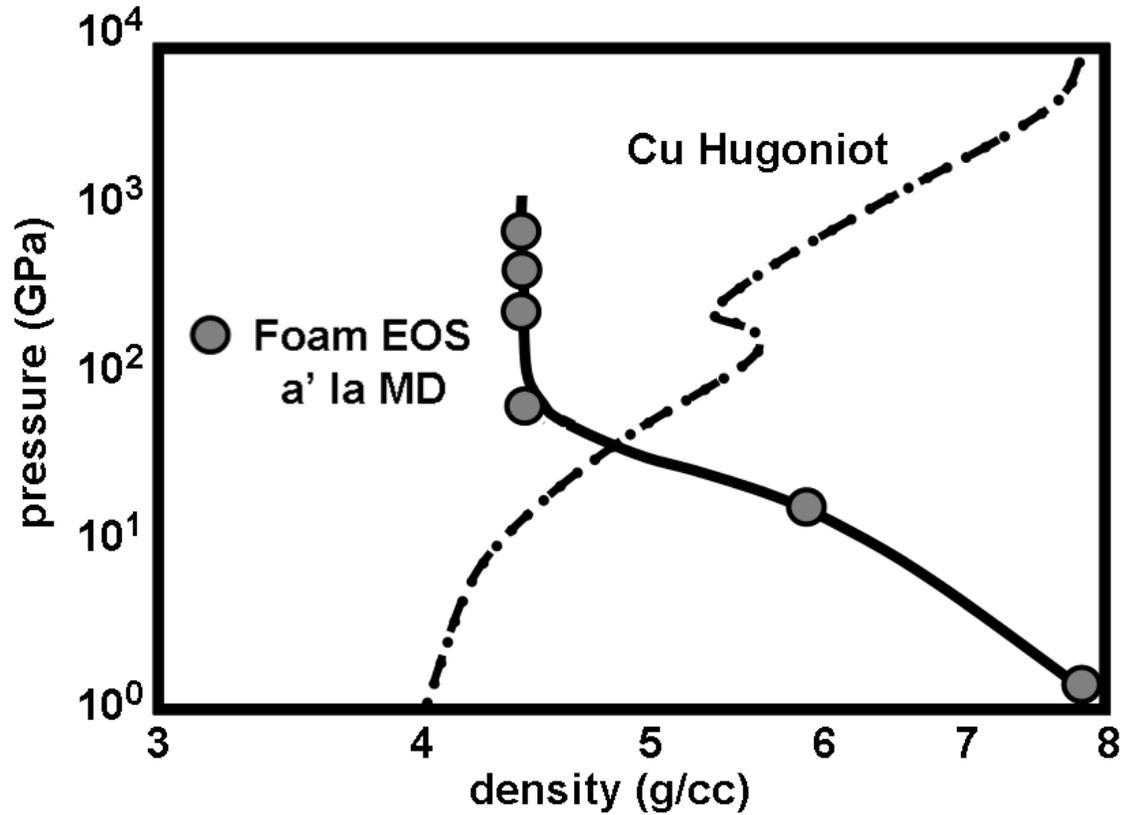

FIG. 11: This study's pressure-density EOS compared to the known copper Hugoniot EOS. The solid line is drawn to guide the eye.

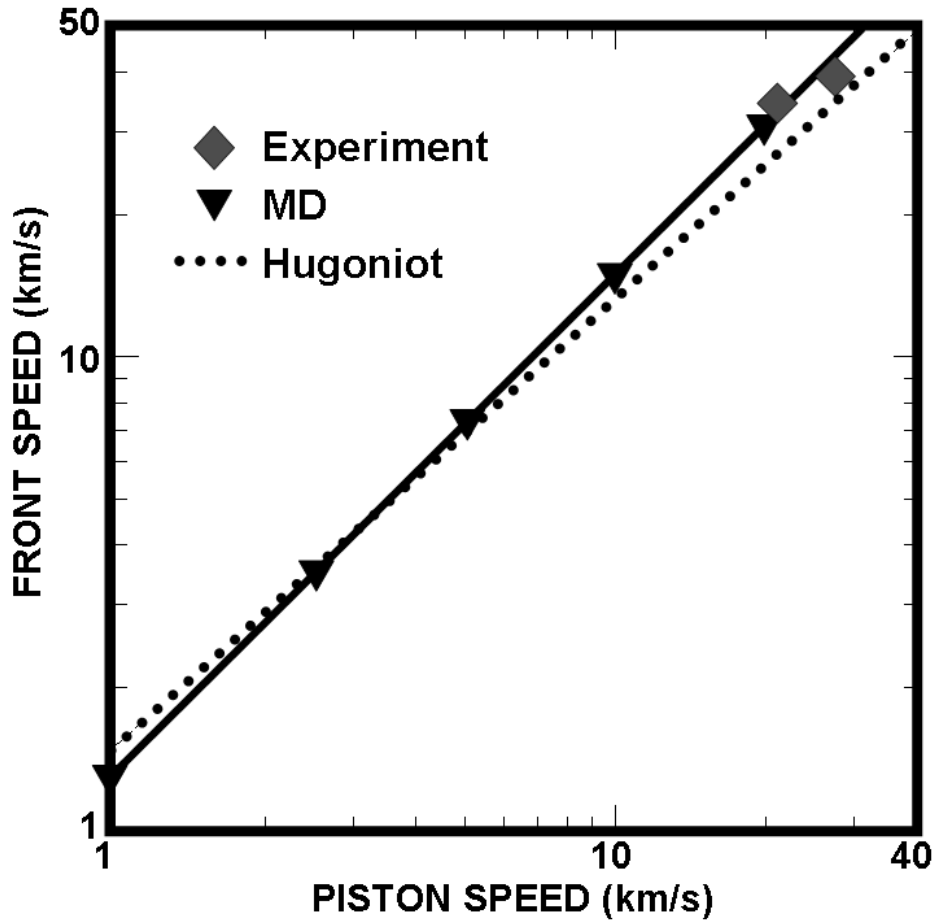

FIG. 12: Comparison of front speed versus piston speed for molecular dynamics simulations (MD), experiment and the copper foam hugoniot. The agreement between all three is good, though MD and experiment is best.

## SUMMARY

This study was driven by the National Ignition Facility (NIF) Energy and Science programs. As emphasized at the beginning, we wished to investigate the generic features of the compaction dynamics of metallic foams as related to current ICF and science interests. A successful atomistic method was implemented for creating a "computer" nanofoam. The microscopic processes in the compaction dynamics showed that a strong compaction wave passing through a foam induces acceleration, crushing, ablation and mixing of the foam material within the porous background. Depending on the piston speed, the ejecta could be a solid front, a liquid, an imperfect vapor, or a hot gas. The simulations agreed with very limited experimental measurements. As should be expected, the simple snowplow model gave an overall accurate macroscopic description of the dynamics. The equation of state of the compacted foam is a direct consequence of the atomistic simulation, though its accuracy depends of a robust description of the atomic behavior over a wide range temperatures and pressures. Our simple interatomic potential did not fulfill that requirement because of the constrained neutrality of the atoms. However, simple model estimates of the ionization were made and are believed to be reasonable. The crushing of the foam is a highly complex evolving state as one moves from the front toward the piston.